\documentclass[11pt,a4paper]{article}
\usepackage[hyperref]{acl2021}
\usepackage{times}
\usepackage{latexsym}
\usepackage{graphicx}
\usepackage{placeins}
\usepackage{subcaption}
\usepackage[capitalize]{cleveref}

\newcommand{\red}[1]{#1} 

\usepackage{microtype}

\aclfinalcopy 

\title{When the Echo Chamber Shatters: Examining the Use of Community-Specific Language Post-Subreddit Ban}

\author{Milo Z. Trujillo \\
  University of Vermont\\
  \texttt{milo.trujillo@uvm.edu} 
  \And
  Samuel F. Rosenblatt \\
  University of Vermont\\
  \texttt{samuel.f.rosenblatt@uvm.edu} 
  \AND
  Guillermo de Anda J\'auregui \\
  National Institute of Genomic Medicine (INMEGEN)\\ 
  Programa de C\'atedras CONACYT para J\'ovenes Investigadores \\
  Universidad Nacional Aut\'onoma de M\'exico \\ 
  \texttt{gdeanda@inmegen.edu.mx} 
  \AND
  Emily Moog \\
  University  of Illinois at Urbana-Champaign \\
  \texttt{emoog2@illinois.edu} 
  \And
  Briane Paul V. Samson \\
  De La Salle University \\
  \texttt{briane.samson@dlsu.edu.ph} 
  \AND
  Laurent H\'ebert-Dufresne\\
  University of Vermont \\
  \texttt{laurent.hebert-dufresne@uvm.edu} 
  \And
  Allison M. Roth \\
  University of Florida \\
  \texttt{amr2264@columbia.edu} 
  }

\begin{document}
\maketitle


\begin{abstract}
Community-level bans are a common tool against groups that enable online harassment and harmful speech. Unfortunately, the efficacy of community bans has only been partially studied and with mixed results. Here, we provide a flexible unsupervised methodology to identify in-group language and track user activity on Reddit both before and after the ban of a community (subreddit). We use a simple word frequency divergence to identify uncommon words overrepresented in a given community, not as a proxy for harmful speech but as a linguistic signature of the community. We apply our method to 15 banned subreddits, and find that community response is heterogeneous between subreddits and between users of a subreddit. Top users were more likely to become less active overall, while random users often reduced use of in-group language without decreasing activity. Finally, we find some evidence that the effectiveness of bans aligns with the content of a community. Users of dark humor communities were largely unaffected by bans while users of communities organized around white supremacy and fascism were the most affected. Altogether, our results show that bans do not affect all groups or users equally, and pave the way to understanding the effect of bans across communities.
\end{abstract}

\section{Introduction}

Online spaces often contain toxic behaviors such as abuse or harmful speech \cite{blackwell2017classification, saleem2017web, jhaver2018online,saleem2018aftermath,habib2019act,ribeiro2020does, de2018hate, sprugnoli2018creating, park2017one, singh2018aggression, lee2018comparative}. Such toxicity may result in platform-wide decreases in user participation and engagement which, combined with external pressure (e.g., bad press), may motivate platform managers to moderate harmful behavior \cite{saleem2018aftermath,habib2019act}. Moreover, the radicalization of individuals through their engagement with toxic online spaces may have real-world consequences, making toxic online communities a cause for broader concern \cite{pizzagate,habib2019act,ribeiro2020does,ribeiro2020auditing}.

Reddit is a social media platform that consists of an ecosystem of different online spaces. As of January 2020, Reddit had over 52 million daily active users organized in over 100,000 communities, known as ``subreddits”, where people gather to discuss common interests or share subject- or format-specific creative content and news \citep{redditfrontpage}. Every post made on Reddit is placed in one distinct subreddit, and every comment on Reddit is associated with an individual post and therefore also associated with a single subreddit. As Reddit continues to gain popularity, moderation of content is becoming increasingly necessary. Content may be moderated in several ways, including: (1) by community voting that results in increased or decreased visibility of specific posts, (2) by subreddit-specific volunteer moderators who may delete posts or ban users that violate the subreddit guidelines, and (3) by platform-wide administrators that may remove posts, users, or entire communities which violate broader site policies. The removal of an entire subreddit is known as a ``subreddit ban,” and does not typically indicate that the users active in the subreddit have been banned.

Given that the ostensible purpose of subreddit bans is to remove subreddits that are in habitual noncompliance with Reddit’s Terms of Service, it is important to understand whether such bans are successful in reducing the offending content. This is especially of interest when the offending content is related to harmful language. Though limited, there is some evidence to suggest that subreddit bans may be effective by certain metrics. Past work has demonstrated that these bans can have both user- and community-level effects \cite{hazel2020communicative, chandrasekharan2017you,saleem2018aftermath,ribeiro2020does,thomas2021behavior,habib2019act}. Several of these studies have suggested that (1) subreddit bans may lead a significant number of users to completely stop using the site, and that (2) following a ban, users that remain on the platform appear to decrease their levels of harmful speech on Reddit \cite{saleem2018aftermath,thomas2021behavior,habib2019act}. \citet{chandrasekharan2017you} also illustrated that postban migrations of users to different subreddits did not result in naive users adopting offensive language related to the banned communities. More work is required to better understand changes in the language of individual users after such bans.

\section{Previous work}

Previous research provides a foundation for investigating the effects of subreddit bans on harmful language and user activity. Detection of offensive content typically takes the form of automated classification. Different machine learning approaches have been applied with \red{varied} success, including but not limited to support vector machines and random forests to convolutional and recurrent neural networks \cite{zhang2019hate, bosco2018overview,gibert2018hate,kshirsagar2018predictive, malmasi2018challenges,pitsilis2018effective,al2019detection, vidgen2020detecting,zimmerman2018improving}. More recently, \citet{garland2020countering} used an ensemble learning algorithm to classify both hate speech and counter speech in a curated collection of German messages on Twitter. Unfortunately, these approaches require labeled sets of speech to train classifiers and therefore risk not transferring from one type of harmful speech (e.g. misogyny) to another (e.g. racism). We therefore aim for a more flexible approach that does not attempt to classify speech directly, but rather identifies language over-represented in harmful groups; i.e., their in-group language. That language is not a signal of, for example, hate speech per se. In fact, any group is likely to have significant in-group language (e.g. hockey communities are more likely to use the word ``slapshot''). However, detection of in-group language can be fully automated in an unsupervised fashion and is tractable.

The majority of past work on bans of harmful communities on Reddit only examined one or two subreddits, often chosen due to notoriety \cite{hazel2020communicative,chandrasekharan2017you,saleem2018aftermath,ribeiro2020does,habib2019act,thomas2021behavior}. Many of these studies focused on the average change in behavior across users and did not consider the factors which may drive inter-individual differences in behavior following a ban \cite{chandrasekharan2017you,saleem2018aftermath,habib2019act}. Different users may respond differently to subreddit bans based on their level of overall activity or community engagement. For example, \citet{ribeiro2020does} found that users that were more active on Reddit prior to a subreddit ban were more likely to migrate to a different platform following a ban. A user’s activity levels prior to a ban also impacted whether activity levels increased or decreased upon migrating to a different platform \cite{ribeiro2020does}. Similarly, \citet{thomas2021behavior} demonstrated that users who were more active in a subreddit prior to a ban were more likely to change their behavior following the banning of that subreddit, but the authors did not investigate the ways in which users changed their behavior. Lastly, \citet{hazel2020communicative} found that a user’s pre-ban activity level within r/alphabaymarket influenced post-ban shifts in communicative activity. 

While we are interested in the effects of moderation on any online community, we study Reddit because the platform is strongly partitioned into sub-communities, and historical data on both subreddits and users are readily available \cite{baumgartner2020pushshift}. Reddit users are regularly active in multiple subreddits concurrently, and unlike other sub-community partitioned platforms like Discord, Slack, or Telegram, we can easily retrieve a user’s activity on \textit{all} sub-communities. This provides an opportunity to understand how the members of a community change their behavior after that community is banned. Furthermore, knowledge of the drivers of inter-individual behavioral differences may permit moderators to monitor the post-ban activity of certain subsets of users more closely than others, which may lead to an increase in the efficacy of platform-wide moderation.

\section{Methodology}

As part of investigating whether different communities respond differently to a subreddit ban, we examine whether top users differ from random users in their change in activity and in-group language usage following community-level interventions. Specifically, we utilize natural language processing to track community activity after a subreddit ban, across 15 subreddits that were banned during the so-called ``Great Ban'' of 2020. We first identified words that had a higher prevalence in these subreddits than on Reddit as a whole prior to a ban. \red{These words do not necessarily correspond to harmful speech but provide a linguistic signature of the community. The strengths and drawbacks of this approach are discussed in the discussion and appendix.} We then compared the frequency of use of community-specific language, as well as the overall activity level of a user (i.e., the number of total comments), 60 days pre- and post-ban for (1) the 100 users that were most active in the banned subreddit 6 months prior to the ban and (2) 1000 randomly sampled non-top users. We predicted that top and random users that remained on the site following a subreddit ban would react differently to the ban, and we anticipated that there would be variation in how different communities responded to a ban.

\subsection{Data Selection}

We selected 15 subreddits banned in June 2020, after Reddit changed their content policies regarding communities that ``incite violence or that promote hate based on identity or vulnerability'' and subsequently banned approximately 2000 subreddits (i.e., ``the Great Ban''). Based on a list of subreddits banned in the Great Ban \footnote{\url{https://www.reddit.com/r/reclassified/comments/fg3608/updated_list_of_all_known_banned_subreddits/}} and an obscured list of subreddits ordered by daily active users \footnote{\url{https://www.redditstatic.com/banned-subreddits-june-2020.txt}}, we examined the subreddits with more than 2000 active daily users and which had not previously become private subreddits. These most-visited subreddits were ``obscured'' by representing all letters except the first two as asterisks, but were de-anonymized as described in the appendix (\cref{sec:appendix_subreddit_deobfuscation}). By selecting highly active subreddits from the Great Ban we can compare many subreddits banned on the same date, and the differences in how their users responded. The list of subreddits we examined is included in \cref{table:subreddit_categorization}.

\subsection{Data Collection}

For each chosen subreddit, we collected all the submissions and comments made during the 182 days before it was banned. This is possible through the Pushshift API\footnote{\url{https://psaw.readthedocs.io/en/latest/}}, which archives Reddit regularly, but may miss a minority of comments if they are deleted (by the author or by moderators) very shortly after they are posted \cite{baumgartner2020pushshift}.
We use this sample of the banned subreddits to identify users from the community and specific language used by the community. To accomplish the former, we examine the ``author'' field of each comment to get a list of users and how many comments they made on the subreddit during the time frame prior to the ban.

To automatically determine in-group vocabulary words for a subreddit, we create a corpus of all text from the comments in a banned subreddit and compare it the baseline corpus to a corpus of 70 million non-bot comments from across all of Reddit during the same time frame. Bot detection is described in \cref{sec:user_selection}. We can gather this cross-site sample by using comment IDs: every Reddit comment has a unique \red{increasing} numeric ID. By taking the comment ID of the first and last comments from our banned sample, and then uniformly sampling all comment IDs between that range and retrieving the associated comments, we can uniformly sample from Reddit as a whole over arbitrary time ranges.

\red{We used this baseline corpus instead of a more standard English corpus because many such standard corpora rely on books, often in the public domain, whose language may be dated and more formal than Reddit comments. These corpora often also lack terms from current events such as sports team names or political figures, which occur frequently across large parts of Reddit.}

\begin{figure*}[htb]
    \centering
    \begin{subfigure}[b]{0.45\textwidth}
        \centering
   \includegraphics[trim=0 0 0 1cm,clip,width=\linewidth]{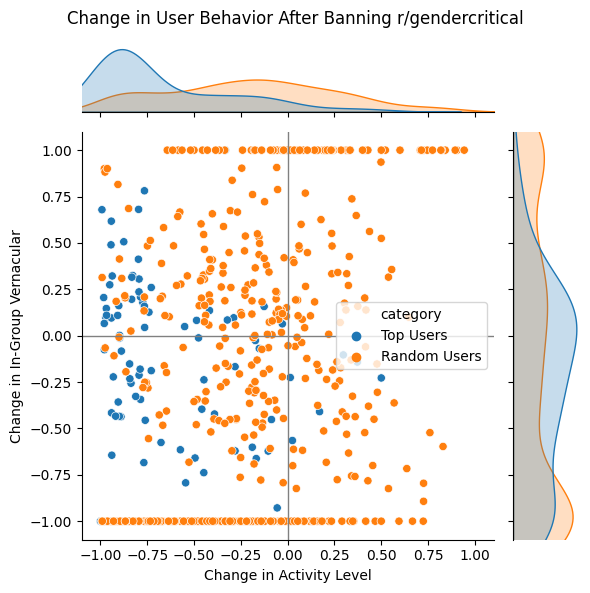}
        \caption{Ban effect on \textit{r/gendercritical} users}
        \label{fig:gendercritical}
    \end{subfigure}
    \begin{subfigure}[b]{0.45\textwidth}
        \centering
    \includegraphics[trim=0 0 0 1cm,clip,width=\linewidth]{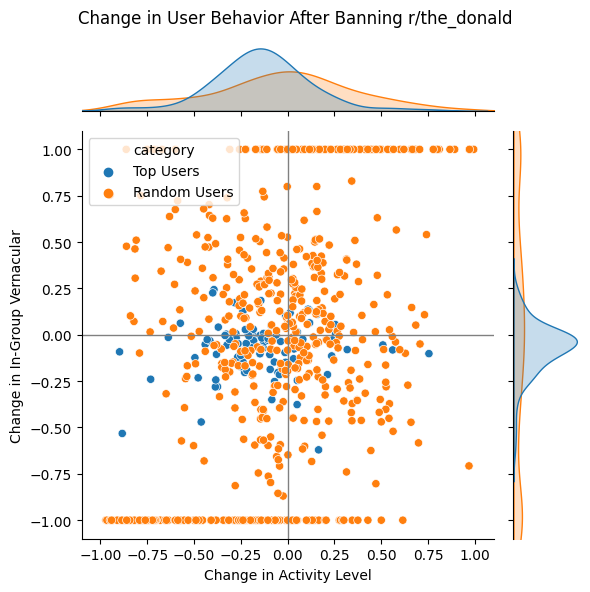}
    \caption{Ban effect on \textit{r/the\_donald} users}
        \label{fig:the_donald}
    \end{subfigure}
    \caption{Example plots comparing user behavior after a subreddit ban. Users from the top 100 and random samples are displayed in terms of their relative change in activity and change in in-group vocabulary usage. Distributions are displayed along each axis for convenience.}
    \label{fig:jointplots}
\end{figure*}

\subsection{Determining In-Group Vocabulary}

We compare word frequencies between the two corpora to identify language that is more prominent in the banned subreddit than in the general sample. Since the two samples are from the same date range on the same platform, this methodology filters out current events and Reddit-specific vocabulary more than we would achieve by comparing to a general English-language corpus like LIWC \cite{tausczik2010psychological}. Rather than comparing relative word occurrence frequency directly, which has pitfalls regarding low-frequency words that may only occur in one corpus, we apply Jensen-Shannon Divergence (JSD) which compares the word frequencies in the two corpora against a mixture text. JSD scores words highly if they appear disproportionately frequently in one corpus, even if they are common in both. \red{For example, JSD identifies ``female'' as a top word in gender-discussion subreddits. Treating ``female'' as in-group vocabulary is undesirable for our specific use-case, where we would prefer to find language specific to the subreddit that is uncommon elsewhere.} Therefore, we remove the top 10,000 most common words in the general corpus from both the general corpus and the subreddit corpus before processing. JSD functionality is provided by the Shifterator software package \cite{gallagher2021generalized}. Based on the resulting JSD scores, we then select the top 100 words in the banned subreddit corpus, and treat this as our final list of in-group vocabulary. \red{We used the top 100 words to maintain consistency with the distinctive vocabulary size used by \citet{chandrasekharan2017you}.} In the appendix, our approach is compared to the Sparse Additive Generative model (SAGE) of \citet{chandrasekharan2017you} to show the additional flexibility of JSD as well as similarity of the results (see \cref{sec:appendix_sage_comparison}).

\subsection{Examining User Behavior} \label{sec:user_selection}

With a list of users from the banned community ranked by comment count and a list of in-group vocabulary, we are able to measure user behavior after the subreddit ban. Since larger subreddits can have tens of thousands to millions of users, we limit ourselves to examining two groups: (1) the 100 most active accounts from a banned subreddit, known as the ``top users'', and (2) a random sample of 1000 non-top users from the subreddit. In forming these lists of top and random users, we skip over accounts from a pre-defined list of automated Reddit bots as well as users that have deleted their accounts and cannot have their post histories retrieved. Additionally, as our focus for this study is users who used in-group language and who continue to use the platform, we omit users that have never used in-group vocabulary pre- or post-ban or who have zero comments post-ban. All forms of user-filtering are discussed further in the appendix (\cref{sec:appendix_omitted_accounts}).

For each user, we download all the comments they made in the 60 days before and after the subreddit ban. We compare the number of comments made before and after the ban to establish a change of activity, on a scale from -1 to 1, with -1 indicating ``100\% of the user’s comments were made prior to the ban'', 0 indicating ``an equal number of comments were made before and after the ban'', and 1 indicating that all of their comments on Reddit were made after the ban. We can similarly track the user’s use of in-group vocabulary on a scale from -1 to 1, for ``100\% of their in-group vocabulary usage was before the ban'' to ``all uses of in-group vocabulary were post-ban''. This is calculated as the fraction of posted words that were in-group vocabulary after the ban, minus the fraction of posted words that we in-group vocabulary before the ban, divided by the sum of the fractions.

$$\frac{r_a - r_b}{r_a + r_b}$$
 Examples of results for individual subreddits are shown in Fig.\ref{fig:jointplots}.


\subsection{Statistical Methods}

\begin{table*}[htb]
\centering
\begin{tabular}{ll}
\hline \textbf{Category} & \textbf{Subreddits} \\ \hline
Dark Jokes         &      darkjokecentral, darkhumorandmemes, imgoingtohellforthis2      \\
Anti-Political         &   consumeproduct, soyboys, wojak    \\
Mainstream Right Wing & the\_donald, thenewright, hatecrimehoaxes \\
Extreme Right Wing & debatealtright, shitneoconssay \\
Uncategorized & ccj2, chapotraphouse, gendercritical, oandaexclusiveforum\\ \hline
\end{tabular}
\caption{Subreddit categorization by qualitative assessment of content}
\label{table:subreddit_categorization}
\end{table*}

\begin{figure*}[htb]
    \centering
    \begin{subfigure}[b]{0.49\textwidth}
        \centering
    \includegraphics[width=\linewidth]{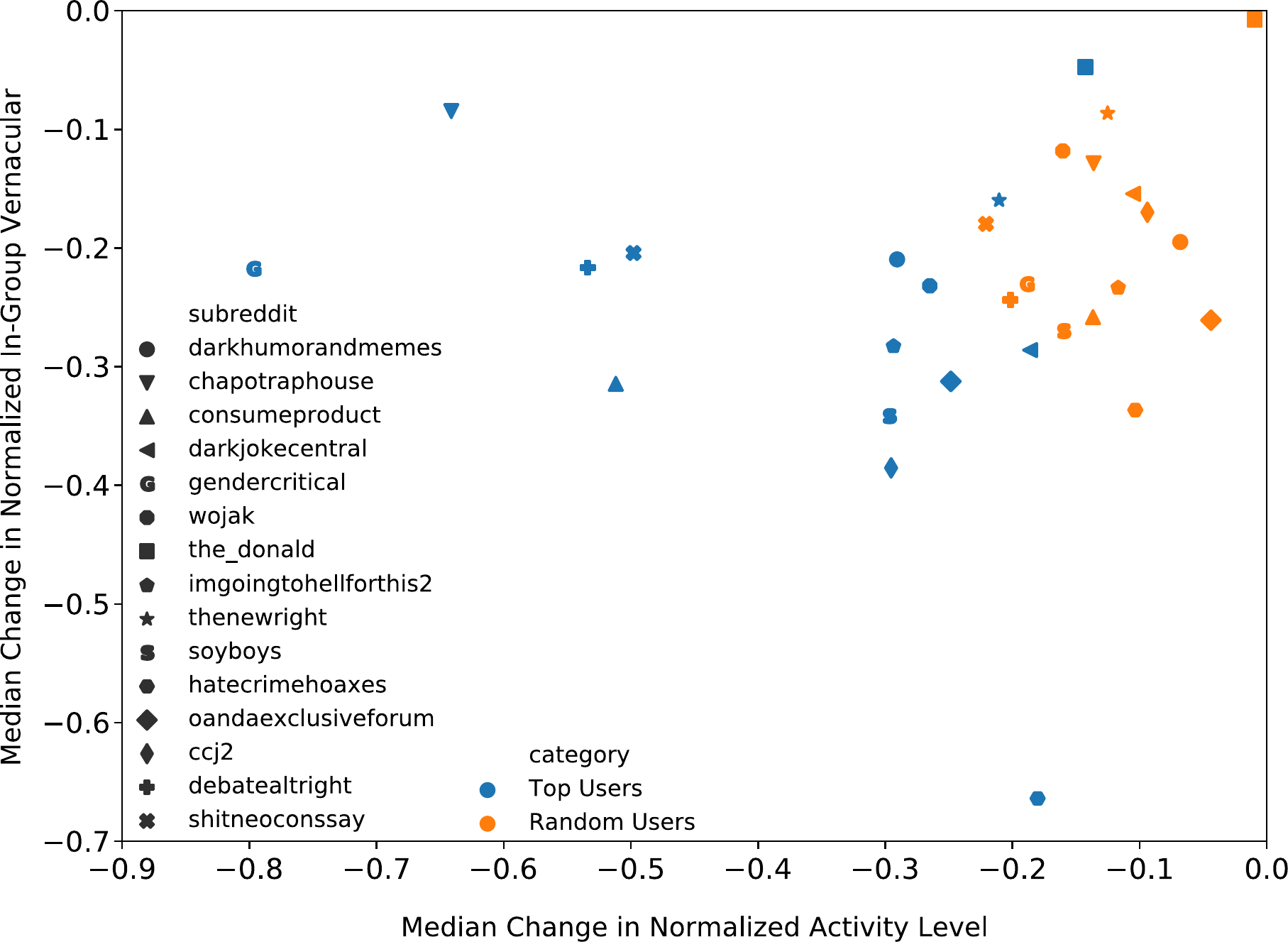}
    \caption{Comparison of top/random users in all 15 subreddits}
        \label{fig:summary_top_bottom}
    \end{subfigure}
    \begin{subfigure}[b]{0.49\textwidth}
        \centering
    \includegraphics[width=\linewidth]{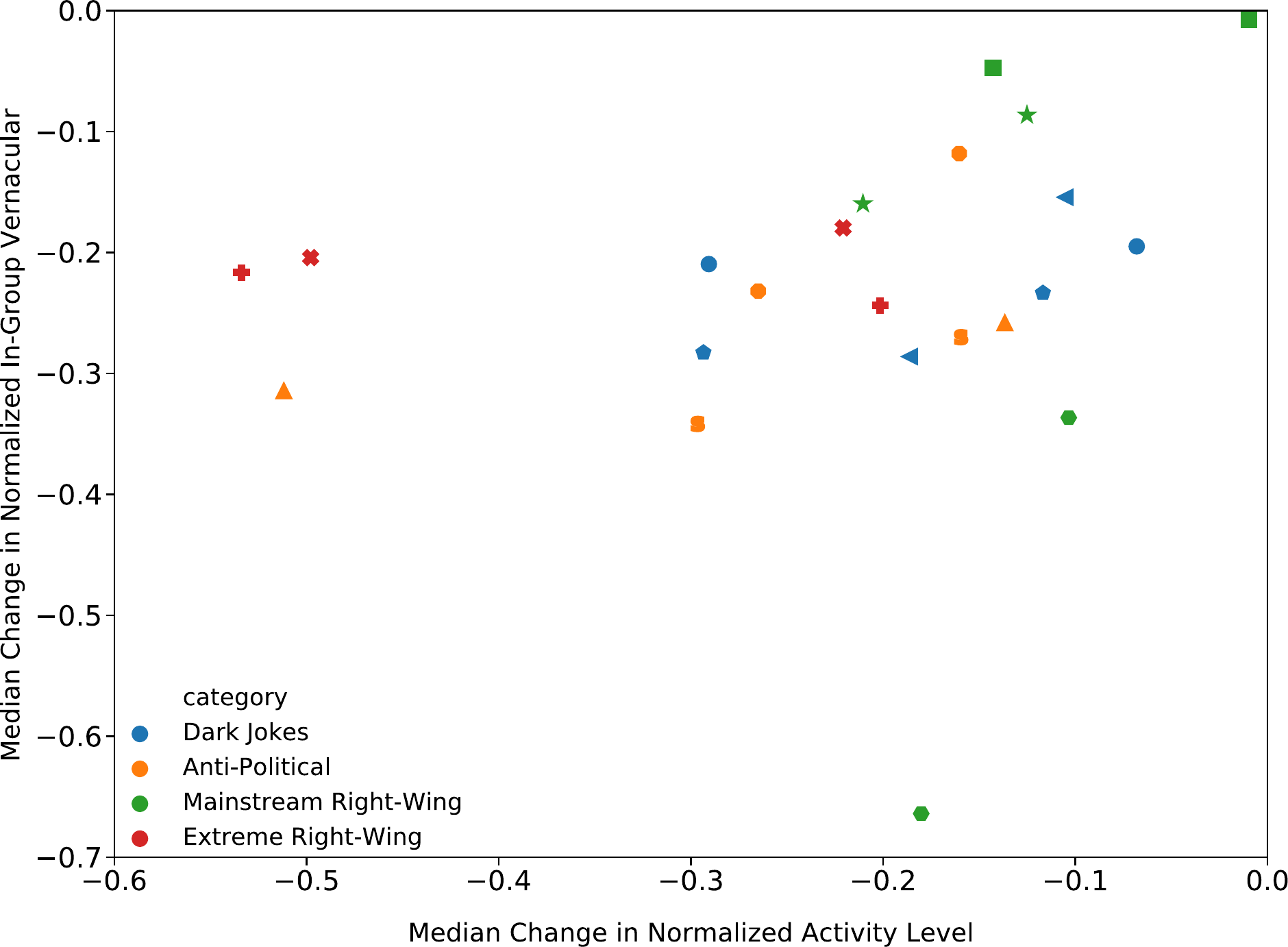}
    \caption{Comparison of top/random users across by categories}
        \label{fig:summary_top_bottom_categories}
    \end{subfigure}
    \caption{Comparison of top and random user behavior changes across fifteen subreddits banned after a change in Reddit content policy in January, 2020. (a) Top users show more significant drop-offs in posting activity after a ban, but have around the same change in in-group vocabulary usage as a uniform sampling of subreddit participants. (b) Ban impact on eleven subreddits categorized by content. Each subreddit appears twice, representing top and random users. Four uncategorized subreddits are excluded from the plot. Trends are summarized in \cref{table:ban_impact_by_category}.}
\end{figure*}

We do not necessarily expect all subreddits to respond to a ban in the same way. From the user data for the 60 days before and after the subreddit's banning, we examined whether there was any difference between subreddits for (1) the proportion of a user's total posts that occurred postban vs preban and (2) the proportion of a user’s total in-group vocabulary that occurred postban vs preban. We also explored whether a user’s engagement in a subreddit (i.e., whether they were a top or random user) influenced either measure. To examine the predictors of the proportion of a user's total posts that occurred postban vs preban, we ran a generalized linear mixed model with a binomial error distribution. This model included the ratio of a user’s posts after the ban to their posts before the ban as the predictor, and subreddit identity and user engagement (i.e., top or random) as fixed effects. To examine the predictors of pre-ban vs post-ban total in-group vocabulary, we ran a second generalized linear mixed model with a binomial error distribution. Its predictor was the ratio of the number of in-group vocabulary words a user used after the ban to the number of in-group vocabulary words that they used before the ban. Subreddit identity and user engagement (i.e., top or random) were fixed effects. For both models, we included user identity (i.e. top or random) as a random effect, since some users were active in more than one of the studied subreddits. Additionally, we used a likelihood ratio test (LRT) to explore whether there was an overall effect of subreddit identity on the proportion of a user's total posts that occurred postban vs preban, and the proportion of a user’s total in-group vocabulary that occurred postban vs preban. In the LRT, we compared each described model to a model without subreddit identity. We also used LRTs to compare models with and without user engagement to assess whether there was an overall effect of user engagement on either measure. 

We performed statistical comparisons in order to understand whether users' vocabulary and activity differed before and after the ban, as well as whether top and random users of a given subreddit experienced similar shifts.

To confirm the shifts displayed in \cref{fig:summary_top_bottom} are meaningful we performed Wilcoxon Signed-Rank tests ($\alpha =FDR=0.05$) on the normalized vocabulary ratios and normalized activity ratios before and after the ban. Except for users of the\_donald (both user types) and the top users of chapotraphouse, these tests decreases in-group vocabulary usage in all subreddit/user-type pairs. The same tests showed the ban had a significant effect on all subreddit/user-type pairs in terms of activity level except for the random users of the\_donald, though these effects were not all decreases.

\begin{figure*}[htb]
    \centering
    \includegraphics[width=0.6\textwidth]{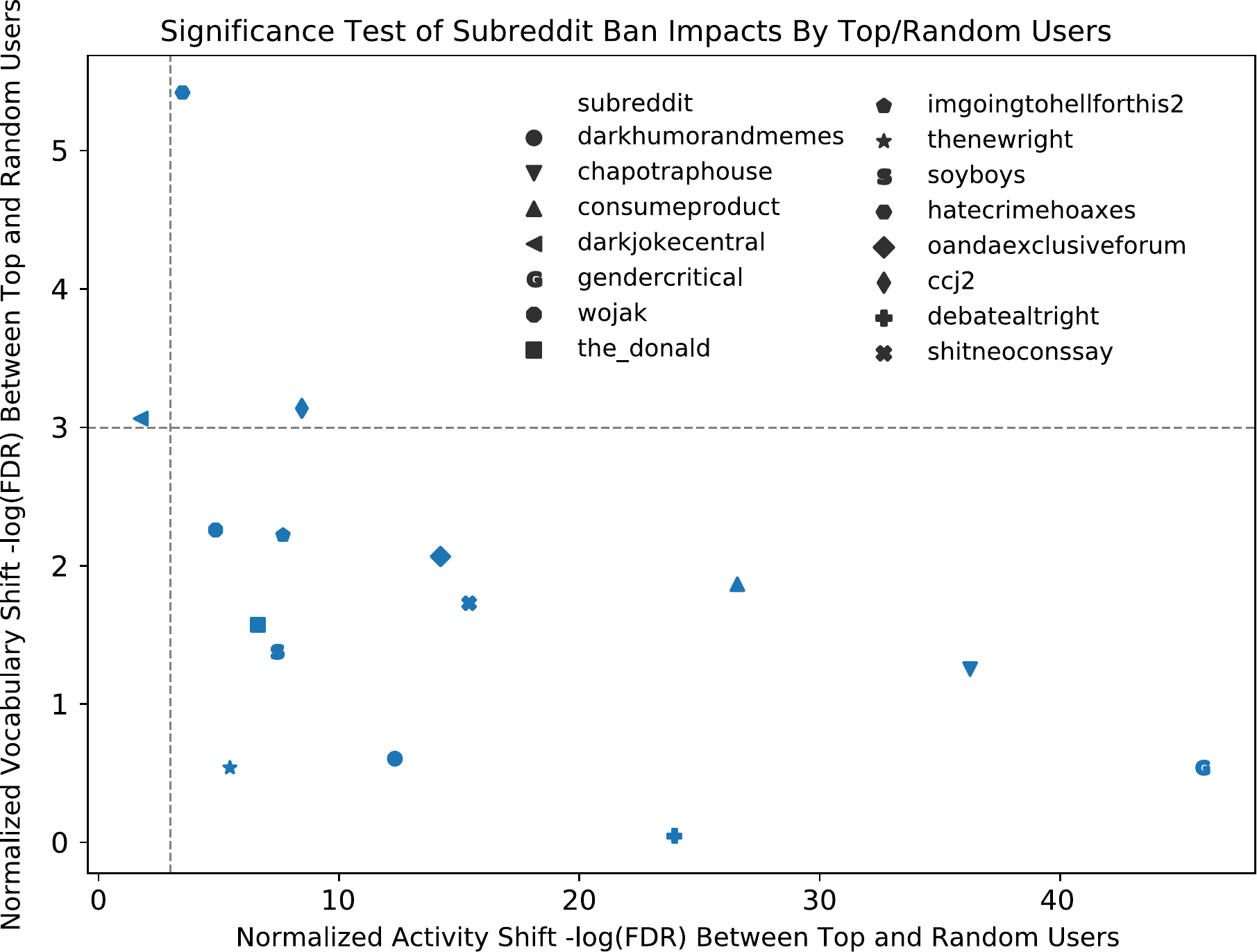}
    \caption{Scatterplot showing differences in activity and vocabulary shifts between top and random users of each subreddit. Each axis shows the statistical significance, expressed as -log(FDR),  of  either activity (x-axis) or vocabulary (y-axis) shift. Dashed lines indicate significance at a threshold of 0.05, such that subreddits with greater values show significant differences between top and random users.}
    \label{fig:cross_subreddit_comparison}
\end{figure*}

We used the Wilcoxon rank sum test to compare the previously defined metrics for vocabulary shift and activity shift between the top and random users within each subreddit. The p-values for each individual comparison at the subreddit level were corrected using false discovery rate (FDR), and are illustrated in \cref{fig:cross_subreddit_comparison}. 

\subsection{Subreddit Categorization}

To better understand our results, we categorized each banned subreddit as ``dark jokes'', ``anti-political'', ``mainstream right wing'', and ``extreme right wing'', as shown in \cref{table:subreddit_categorization}. These categories encompass eleven of our fifteen subreddits, leaving four that are significantly distinct from their peers. Note that the ``uncategorized'' subreddits are not necessarily difficult to classify (for example, r/gendercritical is a trans-exclusionary radical feminist subreddit), but without similar banned subreddits of comparable size we cannot suggest that results for these subreddits are generalizable. While these categories were chosen based on qualitative assessment of each subreddit's content, they are verified by a quantitative comparison of the unique vocabulary of each subreddit available in the appendix. 

\begin{figure*}[htb]
    \centering
    \begin{subfigure}[b]{0.49\textwidth}
        \centering
        \includegraphics[trim=0 0 0 1.2cm,clip,width=1.0\linewidth]{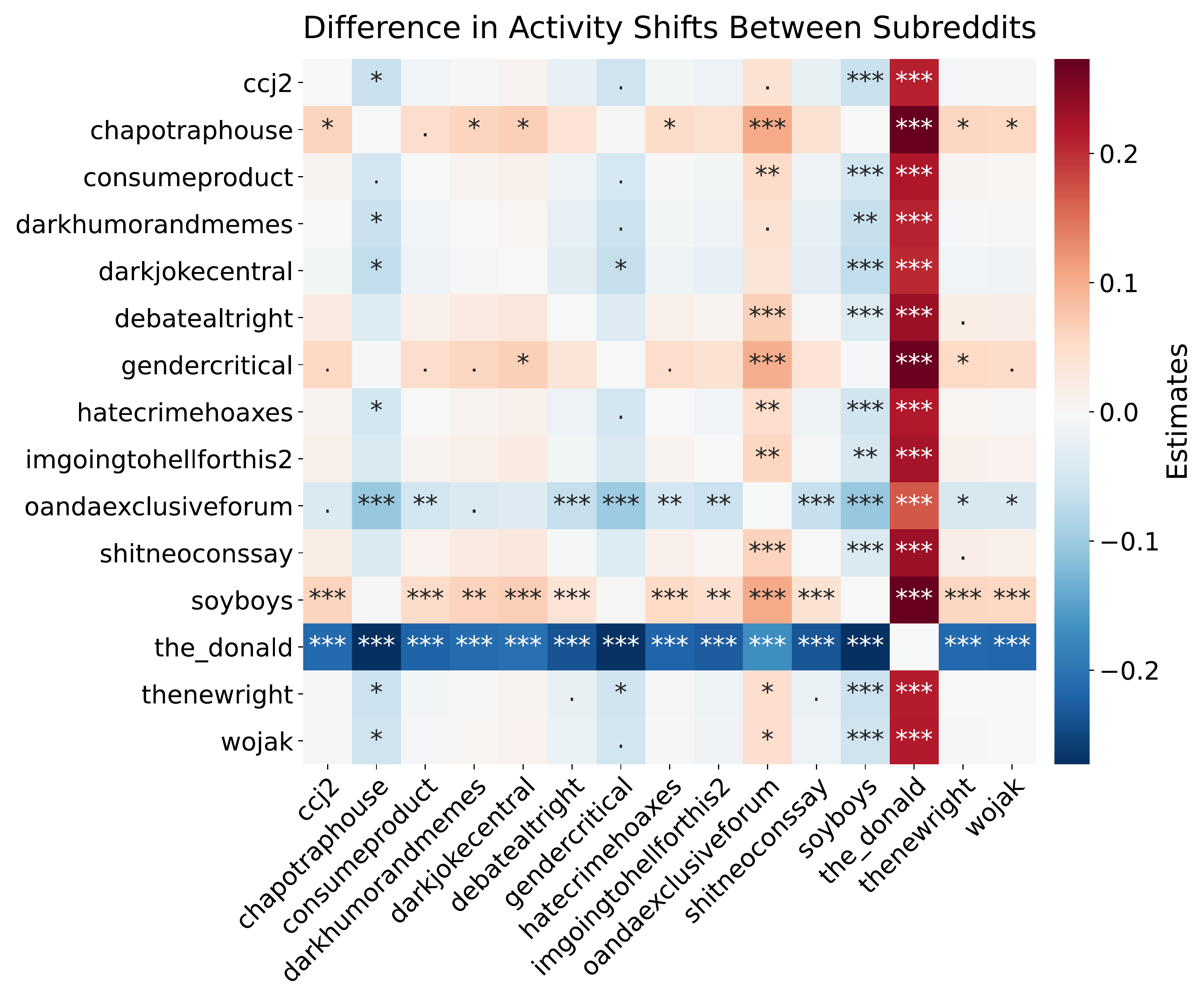}
        \caption{Proportion of Total Posts Post/Pre-ban}
        \label{fig:post_model}
    \end{subfigure}
    \hfill    
    \begin{subfigure}[b]{0.49\textwidth}
        \centering
        \includegraphics[trim=0 0 0 1.2cm,clip,width=1.0\linewidth]{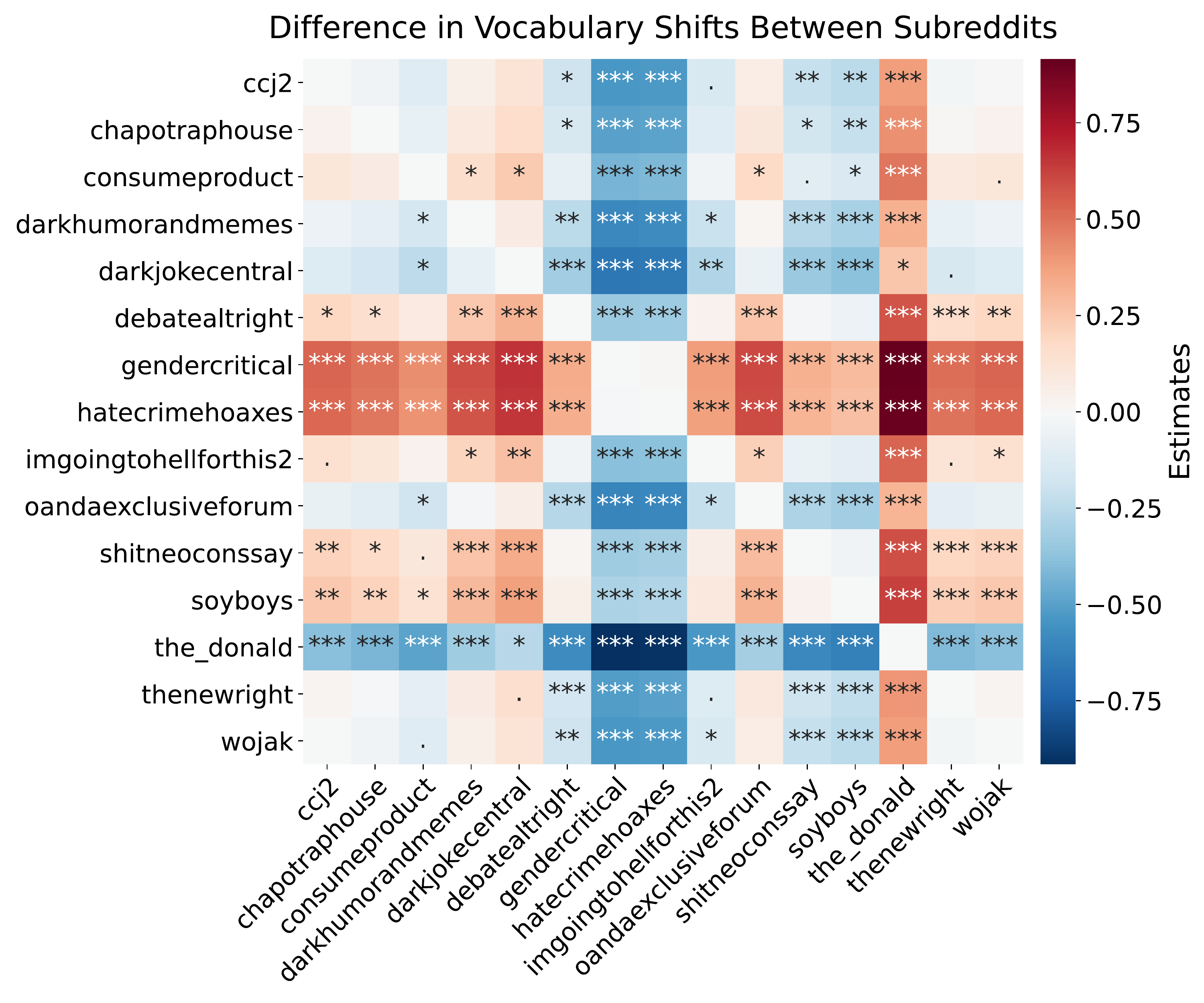}
        \caption{Proportion of Total In-Group Vocabulary Post/Pre-ban}
        \label{fig:vocab_model}
    \end{subfigure}
    \caption{Visualization of GLMM results showing differences between subreddits in postban behavior. For each row, blue cells indicate that the subreddit in a given column had a lower proportion of postban activity/ingroup vocabulary use than the subreddit in that row, while red cells indicate that the subreddit in a given column had a higher proportion of postban activity/ingroup vocabulary use than the subreddit in that row. · indicates p $<$ 0.10. * indicates p $<$ 0.05. ** indicates p $<$ 0.01. *** indicates p $<$ 0.001.}
    \label{fig:heatmaps}
\end{figure*}

\section{Results}

By comparing the median change in activity and vocabulary usage among top and random users, we found a consistent pattern: Top users, for every subreddit studied, decrease their activity more than their peers. This result is important to keep in mind when a uniform sampling of subreddit users post-ban may indicate that a community ban was ineffective. We do not find as consistent a difference between top and random user when looking at vocabulary change; suggesting that while bans may drive harmful users to inactivity, they are less clearly effectual at reforming user behavior. These results are summarized in \cref{fig:summary_top_bottom}. 

To confirm our findings, we tested the statistical significance of differences between top and random distributions for each subreddit, illustrated in \cref{fig:cross_subreddit_comparison}. In all subreddits, there was a significant difference between top and random user changes in either activity shifts, vocabulary shifts, or both. Considering a significance threshold on the false discovery rate, FDR $<$ 0.05, we found two subreddits (r/ccj2 and r/hatecrimehoaxes) that show significant differences in both shifts. The subreddit r/darkjokecentral shows significant differences between top and random users in vocabulary shift, but not activity; whereas the rest of the subreddits show differences in activity but not vocabulary shift between top and random users.

We found that, controlling for user engagement (i.e., whether a user was a top or random user), there was a significant overall effect of subreddit identity on both the proportion of a user's total posts that occurred postban vs preban (LRT, Chi-squared = 133.730, p $<$ 0.001) and the proportion of a user’s total in-group vocabulary that occurred postban vs preban (LRT, Chi-squared = 239.680, p $<$ 0.001). Controlling for subreddit identity, there was also a significant overall effect of user engagement on the proportion of a user's total posts that occurred postban vs preban (LRT, Chi-squared = 23.452, p $<$ 0.001) and the proportion of a user’s total in-group vocabulary that occurred postban (LRT, Chi-squared = 220.020, p $<$ 0.001). Postban posts made up a lower proportion of a user's total posts and postban use of in-group vocabulary made up a lower portion of a user's total in-group vocabulary use for top users compared to random users (\cref{fig:heatmaps}). There were a few subreddits that were significantly different from most or all of the other subreddits. For example, in r/the\_donald, postban posts comprised a higher proportion of a user's total posts, compared to all other subreddits (\cref{fig:post_model}), and postban use of in-group vocabulary comprised a higher portion of a user's total in-group vocabulary use, compared to all other subreddits (\cref{fig:vocab_model}). Postban posts also comprised a higher proportion of a user's total posts in r/oandaexclusiveforum, compared to most other subreddits, while postban posts comprised a lower proportion of a user's total posts in r/soyboys, compared to most other subreddits (\cref{fig:post_model}). The proportion of a user's total in-group vocabulary that occurred postban was lower for both r/gendercritical and r/hatecrimehoaxes, compared to most other subreddits (\cref{fig:vocab_model}).

\begin{table*}[htb]
\centering
\begin{tabular}{lll}
\hline \textbf{Category} & \textbf{Activity Impact} & \textbf{Vocabulary Impact} \\ \hline
Dark Jokes         &      Minimal      & Minimal \\
Anti-Political         &   Top users less active & Decrease among top users  \\
Mainstream Right Wing & Minimal & Inconsistent \\
Extreme Right Wing & All users decrease, especially top users & Minimal \\ \hline
\end{tabular}
\caption{The impact of subreddit bans within each category.}
\label{table:ban_impact_by_category}
\end{table*}

\section{Discussion}

Past work has been quick to conclude that subreddit bans either are \cite{chandrasekharan2017you,saleem2018aftermath,thomas2021behavior} or are not \cite{habib2019act} effective at changing user behavior. We have found that results differ between subreddits and between more and less active users within a subreddit. Since many prior studies on banning efficacy focus on one to two subreddit case studies, these distinctions may not have been apparent in some previous datasets. 

\red{To automatically study a larger number of communities, we tackle the simpler problem of tracking user activity and use of in-group language rather than more subjective harmful language. This approach has strengths and drawbacks. On the one hand, in-group language is easier to automatically identify with little expert knowledge or human intervention, while also including lesser known slang terms or dog whistles that could be harmful. On the other hand, our approach requires a large reference corpus that controls for relevant features of the studied corpus to produce meaningful results. For Reddit, using non-banned subreddits as a baseline corpus allows us to automatically study changes in activity and language around community bans while requiring little expert knowledge on these communities. However, choosing a reference corpus may be more challenging on other platforms without a broader ``mainstream" population (such as alt-tech platforms), with small populations, or without a clear means of sampling the overall population (such as Slack, Discord, and Telegram).}

Our study examines 15 subreddits with over 5000 daily users that were banned simultaneously after a change in Reddit content policy, and our results suggest that subreddit bans impact top and random users differently (in agreement with prior studies such as \citet{hazel2020communicative,ribeiro2020does,thomas2021behavior}) and that community-level banning has a heterogeneous impact across subreddits.

Additionally, we see patterns in subreddit responses to bans that loosely correlate with the type of content the community focused on, summarized in \cref{table:ban_impact_by_category} and illustrated in \cref{fig:summary_top_bottom_categories}. Dark joke subreddits were banned for casual racism, sexism, or other bigotry, do not have as clearly defined in-group language, and were largely unaffected by bans. Users are not more or less active, and use similar language pre and post-ban. Anti-political subreddits, who ridicule most activism and view social progressiveness as performative, were moderately impacted by bans. Top users from these communities became less active after the ban, and randomly sampled users commented using less in-group language. Mainstream right-wing communities show the least consistency in ban response. The most impacted subreddits were extreme political communities that blatantly advocated for white supremacy, anti-multiculturalism, and fascism. These communities saw median top user activity drop to under a third of pre-ban levels, followed by a significant decrease in random user activity, and a modest decrease in in-group vocabulary usage (about -0.2 to -0.3 for all user groups). Since our sample includes only two to four subreddits per category, these trends are not robust but suggest that some pattern might exist within the heterogeneous responses to community-level bans. These results could guide future moderation of online spaces and therefore merit further investigation.

\section{Conclusion}

We have provided a broad investigation of the impact of banning online communities on the activity and in-group vocabulary of the users therein. Our work expands the scope of other studies on this subject, both in terms of the number and types of communities examined. Through this more comprehensive analysis, we have demonstrated heterogeneity in the impact of bans, depending on the type of subreddit and the level of user engagement.  We found that top users generally showed greater reductions in activity and in-group vocabulary usage, compared to random users.  We also found that the efficacy of banning  differs across subreddits, with subreddit content potentially underlying these differences. However, while we provide strong evidence of heterogeneity in ban efficacy, even more comprehensive research must be conducted on a larger group of subreddits in order to fully understand the dynamics behind this heterogeneity.

\section{Future Work}

\red{This study finds heterogeneity in the outcomes of the largest online communities banned on Reddit at the community level and at the individual level. Though we find a clear trend relating outcomes to pre-ban activity level between the top and random users, there are likely other factors at play. Future work could investigate which factors correlate with individual user responses to subreddit bans, including: user demographics (both those directly measurable, such as age of account, and those like gender or country of residence ascertained via tools such as machine learning classifiers), more complex activity metrics (e.g. position of users in interaction networks within the community), and activity in other communities (as measured by number and label of other communities engaged with and level and response of engagement within those communities).}

\red{While we find evidence that community-level responses to bans loosely correlate with the content of the subreddit, our limited sample size of 15 subreddits precludes any thorough quantitative comparisons. Unfortunately, including subreddits with fewer users than the 15 we selected would make community-level statistics less consistent. Were a future study to include large banned subreddits from before or after the “great ban”, identifying the factors and mechanisms that contribute to the differences in subreddit responses would be an important contribution. Potential such factors include: the demographic makeup of the communities, interaction types within the community (potentially measured via network analysis of the comment interaction network of the community), and position in a subreddit-subreddit network of shared users. Studies examining longer-term impacts of community bans would also benefit from considering when some communities attempt to ``rebuild" in a new subreddit, versus integrate into existing subreddits, or rebuild off Reddit entirely.}

\red{However, we believe the most valuable insights may come from embracing more holistic, qualitative methodologies to characterize these banned communities and their responses to moderation. While quantitative metrics indicate heterogeneous community responses, researchers from anthropology and sociology, as well as communications and media studies, may find additional depth in community and user response to censorship. Computational linguists may be able to refine techniques for detecting in-group vocabulary, while linguists and cultural evolution specialists may be best equipped to determine how these vocabularies drift over time. Finally, social computing experts may be in the best position to adapt these multidisciplinary findings to improve platform moderation tools and policies.}

\section{Acknowledgements}
The authors wish to thank the Complex Networks Winter Workshop (CNWW) where the project started, CNWW mentors Daniel B. Larremore, Peter S. Dodds, and Brooke Foucault Welles, as well as Upasana Dutta and Achille Brighton who participated in an early iteration of this work. 

M.Z.T. and L.H.-D. were supported by Google Open Source under the Open-Source Complex Ecosystems And Networks (OCEAN) project. S.F.R. is supported as a Fellow of the National Science Foundation under NRT award DGE-1735316. Any opinions, findings, and conclusions or recommendations expressed in this material are those of the authors and do not necessarily reflect the views of the funders.

\FloatBarrier

\bibliographystyle{acl_natbib}
\bibliography{references}

\clearpage

\section{Appendix}

\subsection{Banned Subreddit De-Obfuscation Process} \label{sec:appendix_subreddit_deobfuscation}

We used a report of the subreddits banned in the ``Great Ban” ranked by daily average users (DAU)~\footnote{\url{https://www.redditstatic.com/banned-subreddits-june-2020.txt}}. The top 20 subreddits with the highest DAU were reported with their names in clear text. The rest of the subreddits had their names obscured, showing only the first two letters and the remaining  characters replaced by asterisks.

To de-obfuscate these, we used the subreddit \textit{r/reclassified} \footnote{\url{https://www.reddit.com/r/reclassified/}}, in which users report banned and quarantined subreddits. We used the Pushshift API to recover posts for the week after the ``Great Ban”, and selected those that had been flagged with the flair \textit{BANNED}.

We then used the following routine to identify the obfuscated banned subreddits from the first list: 

For a given sequence of two initial letters and a given subreddit name length, let $N$ be the number of obscured subreddits with this sequence and name length. Let $M$ be the number of purged subreddits with this initial sequence of letters and length. The $M$ purged subreddits are therefore candidates for the $N$ obscured subreddits.

If $N\geq M$, disambiguate the $N$ obscured subreddits as the $M$ purged subreddits. Any unmatched obscured subreddits are omitted from our analysis.

If $N<M$, manually select the $N$ most-populous subreddits from the $M$ candidate subreddits. Number of commenters was manually researched in the \url{https://reddit.guide/} page for the candidate subreddits.


\subsection{Comparison of Keyword-Selection Methods} \label{sec:appendix_sage_comparison}

The identification of community specific keywords or the identification of hateful speech is an essential part of the pipeline for any kind of analysis on the effect of interventions on online speech. Just as there are numerous methods for the identification of hateful speech \cite{de2018hate, park2017one, singh2018aggression, lee2018comparative}, there are numerous related methods for the identification of community-specific keywords. \citet{chandrasekharan2017you} used a topic modelling framework to identify keywords for their study called the Sparse Additive Generative model (SAGE) which compares ``... the parameters of two logistically-parameterized multinomial models, using a self-tuned regularization parameter to control the tradeoff between frequent and rare terms." The core of this method, the parameter comparison of two logistically-parameterized multinomial models, performs a similar task as our ranking of the contributions of each term to the overall Jensen Shannon Divergence (JSD), and the regularization parameter performs a similar task as our explicit removal of the most common terms in our baseline corpus. As both our methodology and that of \citet{chandrasekharan2017you} perform comparable steps to achieve a comparable outcome, one would expect comparable results. This is somewhat the case when the results are defined for both methods as we can see in the table  \ref{table:subreddit_vocabulary_overlap} below by considering the intersection of terms. However, an important feature of Jensen Shannon Divergence is how it addresses the ``out-of-vocabulary problem" where an instance of a term of any frequency in one corpus has infinitely higher relative frequency than in a compared corpus if that compared corpus does not contain that term. Simplistically, JSD addresses this issue by comparing both corpora to a reference corpus made up of an amalgamation of the two. The SAGE methodology on the other hand, does not have an answer to this problem laid out and so without additional modifications, the SAGE coefficients for such terms that appear in a subreddit of interest but not in a baseline corpus are undefined, and a list of keywords is methodologically impossible to ascertain. As such, we argue that using our JSD-based methodology is more robust to this out-of-vocabulary problem and thus more widely applicable in a variety of settings. Additionally, we view the explicitness of our keyword selection methodology as an advantage compared to the relative ``black box" nature of SAGE.

However, despite the fact that the SAGE-based keyword selection methodology yielded undefined values for a number of the subreddits we studied, given the importance of \citet{chandrasekharan2017you} as foundational to our work, we developed a small extension to the SAGE-based methodology which provides estimates of what the SAGE coefficients would be with a baseline corpus of the entire population of Reddit comments rather than only a sample (note that such a baseline corpus would no longer face this out-of-vocabulary problem as all terms in the subreddit of interest would appear in the population since the subreddit of interest is part of the population). The way these estimates were reached was to use additional known metadata to estimate the counts of all the terms in the baseline corpus as well as the terms in the subreddit of interest which did not appear in the baseline. This was achieved as follows: First, take the frequency counts of each word in the baseline corpus and normalize them to calculate the empirically estimated probability mass function for words in the population of all comments on Reddit for our 6 month timeframe. Second, estimate the number of words on Reddit during this timeframe by taking the exact number of comments on Reddit during this timeframe (calculated by subtracting the first comment ID from this timeframe from the last comment ID from this timeframe) and multiplying this number by the mean number of words per comment in the baseline corpus of 70 million random comments. Third, multiply this estimated number of words on Reddit by the estimated probability mass function for each word to calculate the estimated count of each word in the population rather than the sample. Fourth, add the counts of the out-of-vocabulary terms to these estimated population-sized counts. In the event that those terms appeared only in the subreddit of interest and nowhere else on Reddit during the timeframe examined, this count will be the exact count for that term in the population and it will be at the approximate relative scale when compared to the estimated counts of the other terms in this new estimated population corpus. Using this newly estimated ``population" baseline corpus, we follow the SAGE-based methodology as in  \citet{chandrasekharan2017you} to determine the set of keywords identified by this methodology. Note that in the event that there are no out-of-vocabulary terms, this method simply scales up the frequencies by a constant amount for each term and as a result, reduces exactly to if this extra step had not been performed, but for cases where the out-of-vocabulary problem presents itself, this allows us to gather a list of terms comparable to that methodology.


\begin{table}[htb]
\centering
\small
\begin{tabular}{ll}
\textbf{Subreddit}             & \textbf{Intersection} \\\hline
ccj2                  & 20           \\
chapotraphouse        & 51           \\
consumeproduct        & 61           \\
darkhumorandmemes     & 46           \\
darkjokecentral       & 17           \\
debatealtright        & 35           \\
gendercritical        & 53           \\
hatecrimehoaxes       & 33           \\
imgoingtohellforthis2 & 36           \\
oandaexclusiveforum   & 9            \\
shitneoconssay        & 31           \\
soyboys               & 51           \\
the\_donald           & 56           \\
thenewright           & 57           \\
wojak                 & 34           \\
\textbf{MEAN }                 & \textbf{39.65}       
\end{tabular}
\caption{Number of shared vocabulary words between our JSD-based keyword selection methodology and the SAGE-based methodology }
\label{table:sage_jsd_overlap}
\end{table}

\begin{figure*}[htb]
    \centering
    \includegraphics[width=\linewidth]{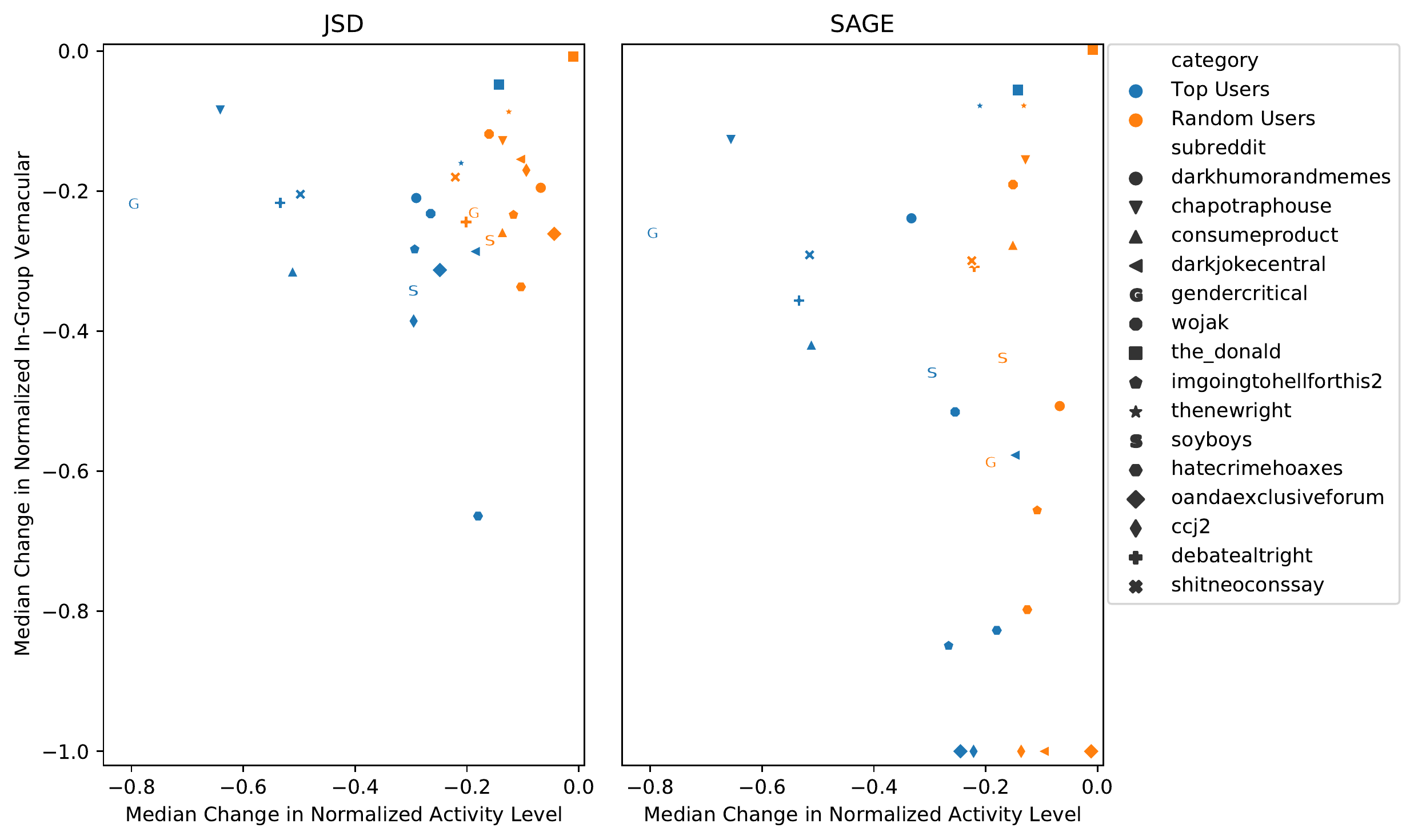}
    \caption{Comparison of top and random user behavior changes under different keyword selection methodology. The subplot on the left corresponds to \ref{fig:summary_top_bottom} in the main text. The differences in activity shift between the two plots are minute and only due to omission of slightly different users for having no in-group vocabulary usage before or after the ban. The relative positions on the vocabulary shift axis remain largely the same except for a wider distribution and several subreddit user-type pairs exhibiting the maximum possible negative shift as the median.}
    \label{fig:SAGE_vs_JSD_summary_top_bottom}
\end{figure*}

Examining figure \ref{fig:SAGE_vs_JSD_summary_top_bottom}, we first notice that for the most part, most subreddit/user-type pairs are in relatively similar positions under the SAGE methodology as under the JSD-based keyword selection, especially when compared relative to each other. \citet{chandrasekharan2017you} found strong negative shifts in in-group vocabulary usage after bans. Upon reproduction of their methodology, we also find stronger negative shifts, including several subreddit/user-type pairs which exhibit a median value of the maximum possible negative vocabulary shift (-1). I.e. the majority of users in these subreddits used at least one SAGE-selected keyword prior to the ban and none thereafter. Examining the data directly, we find that among the subreddit/user-type pairs where this occurred, all five had over half of their users use a SAGE-identified in-group vocabulary word between one and three times only prior to the ban. Additionally, three out of five had a majority use a SAGE-identified in-group vocabulary word one to three times prior to the ban and then zero times after the ban. Under the JSD-based methodology, no subreddit/user-type exhibited behavior where the majority of the users ceased all vocabulary usage after the ban.


The implication that the words chosen by SAGE are not used frequently by a majority of the users of subreddits they are selected from, and are thus not ideally representative, is further supported by the fact that a much larger portion users initially collected had to be omitted due to having zero vocabulary word usage before or after the ban. For the JSD-based methodology, an average of 263 of the initially collected 1000 users were omitted for having never used a single JSD-selected keyword at any time. Under the SAGE-based methodology, this number was 158 users higher on average. I.e. there was a substantially greater portion of users who used no SAGE identified vocabulary words either before or after the ban than users who used no JSD-identified vocabulary words.

The omissions mentioned above are the only cause of differences in activity shift between the the two methodologies. Apart from which users were omitted, the users studied under each methodology were identical and thus had identical activity shifts.

\begin{table*}[bth]
\centering
{
\small
\begin{tabular}{llll}
\hline \textbf{Subreddit} & \textbf{1st Match} & \textbf{2nd Match} & \textbf{3rd Match} \\ \hline
ccj2 & imgoingtohellforthis2 (4) & darkhumorandmemes (3) & chapotraphouse (2) \\ 
chapotraphouse & shitneoconssay (8) & consumeproduct (7) & thenewright (5) \\ 
consumeproduct & wojak (37) & soyboys (37) & shitneoconssay (19) \\ 
darkhumorandmemes & imgoingtohellforthis2 (22) & darkjokecentral (18) & wojak (11) \\ 
darkjokecentral & darkhumorandmemes (18) & imgoingtohellforthis2 (7) & wojak (4) \\ 
debatealtright & shitneoconssay (49) & thenewright (30) & consumeproduct (14) \\ 
gendercritical & darkhumorandmemes (5) & consumeproduct (3) & soyboys (2) \\ 
hatecrimehoaxes & imgoingtohellforthis2 (14) & thenewright (6) & debatealtright (5) \\ 
imgoingtohellforthis2 & darkhumorandmemes (22) & thenewright (16) & soyboys (14) \\ 
oandaexclusiveforum & darkhumorandmemes (4) & wojak (4) & imgoingtohellforthis2 (3) \\ 
shitneoconssay & debatealtright (49) & thenewright (29) & consumeproduct (19) \\ 
soyboys & consumeproduct (37) & wojak (26) & imgoingtohellforthis2 (14) \\ 
the\_donald & thenewright (15) & shitneoconssay (11) & consumeproduct (7) \\ 
thenewright & debatealtright (30) & shitneoconssay (29) & imgoingtohellforthis2 (16) \\ 
wojak & consumeproduct (37) & soyboys (26) & imgoingtohellforthis2 (13)   \\ \hline
\end{tabular}
}
\caption{Comparison of subreddits based on number of shared terms in their respective top 100 in-group vocabulary. These number of shared terms, shown in parenthesis, reinforce qualitative categorization in \cref{table:subreddit_categorization}}
\label{table:subreddit_vocabulary_overlap}
\end{table*}

\subsection{Validation of Subreddit Categories by Vocabulary Overlap} \label{sec:appendix_category_validation}

We initially classified each subreddit by a qualitative assessment of community content. However, we can hypothesize that subreddits with similar focuses are more likely to share in-group vocabulary terms, or conversely, that unrelated subreddits with divergent content are unlikely to share in-group vocabulary. Therefore, if our categorization is accurate, subreddits in each category should share more in-group vocabulary with one another than with other subreddits. This is easily tested, and the results are shown in \cref{table:subreddit_vocabulary_overlap}.

\subsection{Accounts Omitted from Analysis} \label{sec:appendix_omitted_accounts}

In order to limit the analysis to human users and exclude any unobservable or misleading data, we excluded from all parts of the pipeline of this research (from keyword identification to vocabulary shift analysis) any comment which was made by a username in an amassed list of non-human ‘bot’ users. Additionally, we excluded any comment which was made by a user who deleted their account between the time of posting and the time of data ingestion by PushShift, as comments made by these users all present with the indistinguishable username “[deleted].” We used a list of bots curated by \url{botrank.pastimes.eu}, which itself uses its own Reddit bot to scrape comments searching for replies to accounts indicating that the replying user considers the account to be a bot. These comments are a common practice on Reddit and take the form of users indicating their approval or disapproval of an account they perceive to be a bot via the phrases ``Good bot/good bot” and ``Bad bot/bad bot” respectively. The system that populates \url{botrank.pastimes.eu} scrapes from all comments on Reddit at intervals and compiles a list of accounts who have had either “good bot” or “bad bot” replied to them, as well as the number of times this has been done for each such account.  The higher the sum of the counts of ``good bot” and ``bad bot” replies, the more users who have identified the given account as a bot (and are expressing their approval or disapproval of this account). Thus, accounts which have high counts of these replies can be considered as very likely to be bots. As such, we assembled the majority of the list of accounts we excluded from our analysis via identifying each such account in the above mentioned compilation which had over 300 occurrences of users reply either ``good bot” or ``bad bot” to them. This contributed 263 accounts we excluded. Additionally, we manually identified two other accounts below this threshold of 300 occurrences as bots by combing through the data (`darkrepostbot', and `tweettranscriberbot').  With the addition of the `[deleted]’ accounts, this resulted in a total of 266 usernames for which comments were excluded from our analysis, which are included in supplementary material.

Because the focus of our study was users who continued to use the platform and who used in-group language, we omitted users who had zero comments after the ban and users who had zero instances of in-group vocabulary usage before or after the ban. No top users fell into either of these categories as they all used in group language either before or after the ban and all made at least one comment after the ban. The breakdown of how many users this final sequence of omissions results in amongst the random users, broken down as subreddit:(number users omitted for having zero postban comments, number users omitted for having no in-group vocabulary usage), is as follows: 
oandaexclusiveforum: (171, 239); ccj2: (174, 264); darkjokecentral: (132, 468); darkhumorandmemes: (146, 477); shitneoconssay:(223, 119) ; imgoingtohellforthis2:( 141, 358); consumeproduct:( 147, 292); the\_donald:( 94, 332); debatealtright:( 257, 118); gendercritical: (207, 278); chapotraphouse:( 108, 222); soyboys:( 203 , 214); hatecrimehoaxes:( 141, 113 ); thenewright:( 128, 190); wojak:(137, 257).

\subsection{Software and Data}
\label{sec:software_and_data}
Software is available for review through anonymous figshare\footnote{\url{https://figshare.com/s/a8f250ed3edfecaa5de3}}, to be published via GitHub. Analysis data included in supplementary material.


\end{document}